\documentstyle[preprint,epsfig,aps,amsmath,floats,tighten]{revtex} 

\def\met{\mbox{${\hbox{$E$\kern-0.6em\lower-.1ex\hbox{/}}}_T~$}} 
\def\D0{D\O}                            
\begin{document}
\lefthyphenmin=2
\righthyphenmin=3  
\setlength{\unitlength}{1.0mm}

%
%
\title{
Inclusive Jet Cross Sections \\in $\overline{p}p$ Collisions at
$\sqrt{s}$= 630 and 1800~GeV
}

\author{\centerline{The D\O\ Collaboration
  \thanks{Submitted to the {\it International Europhysics Conference} on
         {\it High Energy Physics}, {\it EPS-HEP99},
          \hfill\break
           15 -- 21 July, 1999, Tampere, Finland.}}}
\address{
\centerline{Fermi National Accelerator Laboratory, Batavia, Illinois 60510}
}

%
%
\date{\today}

\maketitle

%
%
\begin{abstract}
We have made a precise measurement of the inclusive jet cross section 
at $\sqrt{s}=1800$~GeV. The result is based on an integrated luminosity 
of 92 pb$^{-1}$ collected at the Fermilab Tevatron $\overline{p}p$ 
Collider with the D\O\ detector. The measurement is reported as a function 
of jet transverse energy ($60$~GeV $\leq E_{T} < $ $550$~GeV), and in 
the pseudorapidity intervals $|\eta|\leq0.5$ and $0.1\leq|\eta|\leq0.7$.
A preliminary measurement of the pseudorapidity dependence of inclusive 
jet production ($|\eta|\leq1.5$) is also discussed. The results are in 
good agreement with predictions from next--to--leading order (NLO) quantum 
chromodynamics (QCD). D\O\ has also determined the ratio of jet cross 
sections at $\sqrt{s}=630$~GeV and $\sqrt{s}=1800$~GeV ($|\eta|\leq0.5$). 
This preliminary measurement differs from NLO QCD predictions.
\end{abstract}

\newpage
\begin{center}
%
B.~Abbott,$^{45}$                                                             
M.~Abolins,$^{42}$                                                            
V.~Abramov,$^{18}$                                                            
B.S.~Acharya,$^{11}$                                                          
I.~Adam,$^{44}$                                                               
D.L.~Adams,$^{54}$                                                            
M.~Adams,$^{28}$                                                              
S.~Ahn,$^{27}$                                                                
V.~Akimov,$^{16}$                                                             
G.A.~Alves,$^{2}$
% ...
\\                                                                            
\vskip 0.30cm                                                                 
\centerline{(D\O\ Collaboration)}                                             
\vskip 0.30cm                                                                 
\centerline{$^{1}$Universidad de Buenos Aires, Buenos Aires, Argentina}       
\centerline{$^{2}$LAFEX, Centro Brasileiro de Pesquisas F{\'\i}sicas,         
                  Rio de Janeiro, Brazil}                                     
\centerline{$^{3}$Universidade do Estado do Rio de Janeiro,                   
                  Rio de Janeiro, Brazil}                                     
\centerline{$^{4}$Institute of High Energy Physics, Beijing,                  
                  People's Republic of China}                                 
\centerline{$^{5}$Universidad de los Andes, Bogot\'{a}, Colombia}             
\centerline{$^{6}$Universidad San Francisco de Quito, Quito, Ecuador}         
\centerline{$^{7}$Institut des Sciences Nucl\'eaires, IN2P3-CNRS,             
                  Universite de Grenoble 1, Grenoble, France}                 
\centerline{$^{8}$DAPNIA/Service de Physique des Particules, CEA, Saclay,     
                  France}                                                     
\centerline{$^{9}$Panjab University, Chandigarh, India}                       
\centerline{$^{10}$Delhi University, Delhi, India}                            
\centerline{$^{11}$Tata Institute of Fundamental Research, Mumbai, India}     
\centerline{$^{12}$Kyungsung University, Pusan, Korea}                        
\centerline{$^{13}$Seoul National University, Seoul, Korea}                   
\centerline{$^{14}$CINVESTAV, Mexico City, Mexico}                            
\centerline{$^{15}$Institute of Nuclear Physics, Krak\'ow, Poland}            
\centerline{$^{16}$Institute for Theoretical and Experimental Physics,        
                   Moscow, Russia}                                            
\centerline{$^{17}$Moscow State University, Moscow, Russia}                   
\centerline{$^{18}$Institute for High Energy Physics, Protvino, Russia}       
\centerline{$^{19}$Lancaster University, Lancaster, United Kingdom}           
\centerline{$^{20}$University of Arizona, Tucson, Arizona 85721}              
\centerline{$^{21}$Lawrence Berkeley National Laboratory and University of    
                   California, Berkeley, California 94720}                    
\centerline{$^{22}$University of California, Davis, California 95616}         
\centerline{$^{23}$University of California, Irvine, California 92697}        
\centerline{$^{24}$University of California, Riverside, California 92521}     
\centerline{$^{25}$Florida State University, Tallahassee, Florida 32306}      
\centerline{$^{26}$University of Hawaii, Honolulu, Hawaii 96822}              
\centerline{$^{27}$Fermi National Accelerator Laboratory, Batavia,            
                   Illinois 60510}                                            
\centerline{$^{28}$University of Illinois at Chicago, Chicago,                
                   Illinois 60607}                                            
\centerline{$^{29}$Northern Illinois University, DeKalb, Illinois 60115}      
\centerline{$^{30}$Northwestern University, Evanston, Illinois 60208}         
\centerline{$^{31}$Indiana University, Bloomington, Indiana 47405}            
\centerline{$^{32}$University of Notre Dame, Notre Dame, Indiana 46556}       
\centerline{$^{33}$Purdue University, West Lafayette, Indiana 47907}          
\centerline{$^{34}$Iowa State University, Ames, Iowa 50011}                   
\centerline{$^{35}$University of Kansas, Lawrence, Kansas 66045}              
\centerline{$^{36}$Kansas State University, Manhattan, Kansas 66506}          
\centerline{$^{37}$Louisiana Tech University, Ruston, Louisiana 71272}        
\centerline{$^{38}$University of Maryland, College Park, Maryland 20742}      
\centerline{$^{39}$Boston University, Boston, Massachusetts 02215}            
\centerline{$^{40}$Northeastern University, Boston, Massachusetts 02115}      
\centerline{$^{41}$University of Michigan, Ann Arbor, Michigan 48109}         
\centerline{$^{42}$Michigan State University, East Lansing, Michigan 48824}   
\centerline{$^{43}$University of Nebraska, Lincoln, Nebraska 68588}           
\centerline{$^{44}$Columbia University, New York, New York 10027}             
\centerline{$^{45}$New York University, New York, New York 10003}             
\centerline{$^{46}$University of Rochester, Rochester, New York 14627}        
\centerline{$^{47}$State University of New York, Stony Brook,                 
                   New York 11794}                                            
\centerline{$^{48}$Brookhaven National Laboratory, Upton, New York 11973}     
\centerline{$^{49}$Langston University, Langston, Oklahoma 73050}             
\centerline{$^{50}$University of Oklahoma, Norman, Oklahoma 73019}            
\centerline{$^{51}$Brown University, Providence, Rhode Island 02912}          
\centerline{$^{52}$University of Texas, Arlington, Texas 76019}               
\centerline{$^{53}$Texas A\&M University, College Station, Texas 77843}       
\centerline{$^{54}$Rice University, Houston, Texas 77005}                     

\end{center}

\vfill\eject

\section{Introduction}

Within the framework of quantum chromodynamics (QCD), inelastic scattering 
between a proton and antiproton is described as a hard collision between 
their constituents (partons). After the 
collision, the outgoing partons manifest themselves as 
localized streams of particles or ``jets''. 
Predictions for the inclusive jet cross section have 
improved in the early nineties with next-to-leading order (NLO) perturbative
QCD calculations \cite{theory} and new, accurately measured parton density 
functions (pdf)\cite{pdfs}.

The D\O\ Collaboration has recently measured and published~\cite{d0paper}
the cross section 
for the production of jets as a function of the jet energy transverse to 
the incident beams, $E_{T}$. The measurement is based on an integrated 
luminosity of about 92~pb$^{-1}$ of $\overline{p}p$ hard collisions collected 
with the D\O\ Detector~\cite{D0detector} at the Fermilab Tevatron Collider.
This result allows a stringent test of QCD, with a total uncertainty
substantially reduced relative to previous results~\cite{old,nCDF}.  
We have also measured the ratio of jet cross sections at two center-of-mass
energies: $630$ (based on an integrated luminosity of about $0.537$ pb$^{-1}$)
and $1800$~GeV. Experimental and theoretical uncertainties
are significantly reduced in the ratio. This is due to the large 
correlation in the errors of the two cross section measurements,
and the suppression of the sensitivity to parton distribution functions (pdf) 
in the prediction. The ratio of cross sections thus provides a stronger
test of the matrix element portion of the calculation than a single cross
section measurement alone. 
Previous measurements of cross section ratios have been performed with 
smaller data sets by the UA2 and CDF~\cite{ratio} experiments.

\section{Jet Reconstruction and Data Selection}

Jets are reconstructed using an iterative jet cone algorithm with a fixed 
cone radius of $\mathcal{R} =$ $0.7$ in $\eta$--$\phi$ space, (pseudorapidity 
is defined as $\eta = -{\rm ln}[{\rm tan}\frac{\theta}{2}]$)~\cite{d0the}.  
The offline data selection procedure, which eliminates background caused by
electrons, photons, noise, or cosmic rays, follows the methods described in
Refs.~\cite{dan,krane}.

\section{Energy Corrections}

The jet energy scale correction, described in \cite{escale}, removes
instrumentation effects associated with calorimeter response, showering, 
and noise, as well as the contribution from spectator partons (underlying 
event).
The energy scale corrects jets from their reconstructed $E_{T}$ to their 
``true'' $E_{T}$ on average. An unsmearing correction is applied later to 
remove the effect of a finite $E_{T}$ resolution~\cite{d0paper}.

\section{The Inclusive Jet Cross Section}

The resulting inclusive double differential jet cross sections, 
$\langle d^2\sigma / (dE_{T} d\eta) \rangle$, 
for $|\eta|\leq0.5$ and $0.1\leq|\eta|\leq 0.7$ (the second region for 
comparison to Ref.~\cite{nCDF}), are compared with a NLO QCD theoretical 
prediction~\cite{theory}.
Discussions on the different choices in the theoretical calculation: pdfs, 
renormalization and factorization scales ($\mu$), and clustering algorithm
parameter ($R_{sep}$) can be found in Refs.~\cite{d0the}.  

Figure~1 shows the ratios $(D-T)/T$ for the data ($D$) and
{\small JETRAD} NLO theoretical ($T$) predictions based on the CTEQ3M, CTEQ4M
and MRST pdf's [4,5] for $|\eta| \leq 0.5$.  
(The tabulated data for both $|\eta| \leq 0.5$ and $0.1 \leq |\eta| \leq 0.7$
measurements can be found in Ref.~\cite{matrix}.)

The predictions are in good quantitative agreement with the data, as
verified with a $\chi^{2} = \sum_{i,j} (D_{i}-T_{i}) (C^{-1})_{ij} (D_{j}-T_{j})$
test, which incorporates the uncertainty covariance matrix $C$. Here $D_{i}$ 
and $T_{i}$ represent the $i$-th data and theory points, respectively. The 
overall systematic uncertainty is largely correlated.

Table~\ref{tab:table2} lists $\chi^{2}$ values for several {\small JETRAD}
predictions using various parton distribution functions~\cite{pdfs}.
The predictions describe both the $|\eta| \leq 0.5$ and 
$0.1 \leq |\eta| \leq 0.7$ cross section very well.
The measurement by D\O\ and CDF are also in good quantitative agreement 
within their systematic uncertainties~\cite{d0paper}.

\vskip0.4cm
\begin{table}[htbp]
\caption{ $\chi^{2}$ comparisons between {\small JETRAD} and $|\eta| \leq 0.5 $
and $0.1 \leq |\eta| \leq 0.7 $ data for $\mu = 0.5E_{T}^{\rm max}$,
$\cal{R}_{\rm{sep}}$=$1.3\cal{R}$, and various
pdfs.  There are 24 degrees of freedom. }
\vskip0.2cm
\begin{tabular}{ccc}
pdf &  $|\eta| \leq 0.5 $ &  $0.1 \leq |\eta| \leq 0.7 $ \\ \hline
  CTEQ3M     &  23.9     &  28.4    \\
  CTEQ4M     &  17.6     &  23.3    \\
  CTEQ4HJ    &  15.7     &  20.5    \\
  MRSA\'     &  20.0     &  27.8    \\
  MRST       &  17.0     &  19.5    \\
\end{tabular}
\label{tab:table2}
\end{table}


\vskip0.4cm
\begin{figure}[htb] \centering
  \label{fig:incjet_prl}
  \begin{picture}(155,70)

    \put(35,6){\begin{picture}(90,70)
                 \epsfxsize=9.0cm
                 \epsfbox{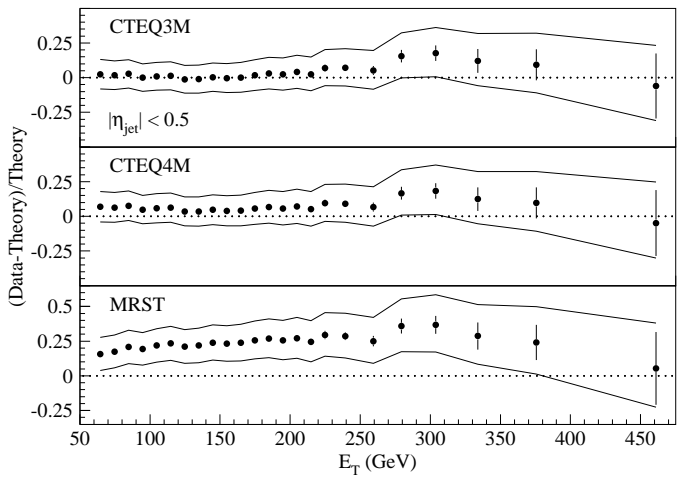}
               \end{picture}}

  \end{picture}
  \caption{ The difference between data and {\small JETRAD} QCD 
	       predictions normalized to predictions. The bands are 
	       the total experimental uncertainty.}
\end{figure}

\section{$\eta$ Dependence of the Inclusive Jet Cross Section}

D\O\ has made a preliminary measurement of the pseudorapidity dependence 
of the inclusive jet cross section. Figure~2 shows the ratios $(D-T)/T$ for 
the data ($D$) and {\small JETRAD} NLO theoretical ($T$) predictions using 
the CTEQ3M pdf set for $0.5\leq|\eta|<1.0$ and $1.0\leq|\eta|<1.5$. The 
measurements and the predictions are in good qualitative agreement. The
pseudorapidity reach of this measurement is currently being extended to 
$\eta=3.0$ and the detailed error analysis is being completed. 

\begin{figure}[htb] \centering
  \label{fig:forward}
  \begin{picture}(155,87)

    \put(35,3){\begin{picture}(85,87)
                 \epsfxsize=8.5cm
                 \epsfbox{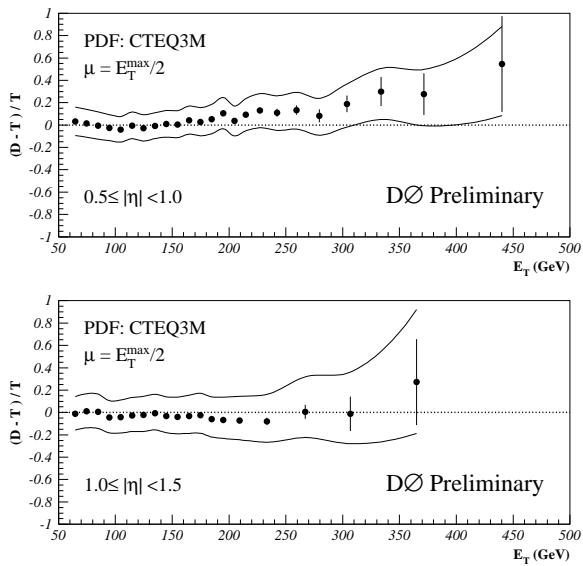}
               \end{picture}}

  \end{picture}
  \caption{ Pseudorapidity dependence of the inclusive jet cross section 
            ($0.5\leq|\eta|<1.0$ and $1.0\leq|\eta|<1.5$). Comparison between 
 	    data and NLO QCD predictions. The bands are the total systematic 
	    uncertainty in the experiment.}
\end{figure}

\section{Ratio of Scale Invariant Jet Cross Sections}

A simple parton model would predict a jet cross section that scales with
center-of-mass energy. In this scenario, $E_{T}^{4} \cdot E 
\frac{d^{3}\sigma}{dp^{3}}$, plotted as a function of jet $x_{T}\equiv
\frac{2 \, E_{T}}{\sqrt{s}}$, would remain constant with respect to the
center-of-mass energy. Figure~3 shows the D\O\ measurement
of $E_{T}^{4} \cdot E \frac{d^{3}\sigma}{dp^{3}}$ (stars) compared to
{\small JETRAD} predictions (lines). There is poor agreement between data
and NLO QCD calculations using the same $\mu$ in the numerator and
the denominator (probability of agreement not greater than 10\%). 
The agreement improves for predictions with different $\mu$ at the two
center-of-mass energies.

\vskip0.3cm

In conclusion, we have made precise measurements of jet production
cross sections. At $\sqrt{s}$=1800~GeV, there is good agreement between
the measurements and the data.
The ratio of cross sections at $\sqrt{s}$=1800 and 630~GeV, however,
differs from NLO QCD predictions, unless different renormalization scales
are introduced for the two center-of-mass energies.


\begin{figure}[htb] \centering
  \label{fig:ratmu}
  \begin{picture}(155,87)

    \put(35,4.5){\begin{picture}(85,87)
                 \epsfxsize=8.5cm
                 \epsfbox{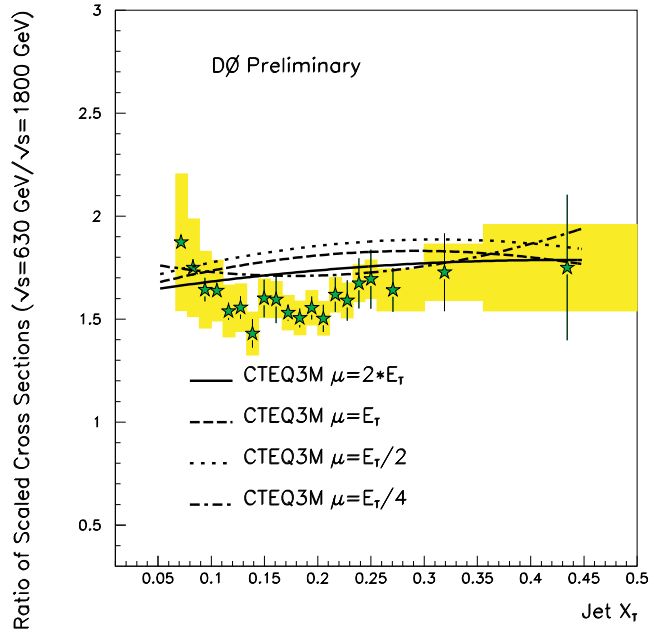}
               \end{picture}}

  \end{picture}
  \caption{ The ratio of scale invariant jet cross sections.
	    The stars are the D\O\ data, the band is the systematic 
	    uncertainty, and the lines are the NLO QCD predictions.}
\end{figure}


\section{Acknowledgements}

%
We thank the Fermilab and collaborating institution staffs for
contributions to this work and acknowledge support from the 
Department of Energy and National Science Foundation (USA),  
Commissariat  \` a L'Energie Atomique (France), 
Ministry for Science and Technology and Ministry for Atomic 
   Energy (Russia),
CAPES and CNPq (Brazil),
Departments of Atomic Energy and Science and Education (India),
Colciencias (Colombia),
CONACyT (Mexico),
Ministry of Education and KOSEF (Korea),
and CONICET and UBACyT (Argentina).
%


\end{document}